%% file: arxiv.tex
\documentclass{article}

\usepackage{arxiv}

\usepackage[utf8]{inputenc} 
\usepackage[T1]{fontenc}    
\usepackage{hyperref}       
\usepackage{url}            
\usepackage{booktabs}       
\usepackage{amsfonts}       
\usepackage{nicefrac}       
\usepackage{microtype}      
\usepackage{lipsum}		
\usepackage{graphicx}
\usepackage{doi}
\usepackage{bbm}
\usepackage{amssymb}
\usepackage{amsmath}
\usepackage{hyperref}

\usepackage{algorithm}
\usepackage{algorithmic}

\title{A Reinforcement Learning Framework for \\ Online Speaker Diarization}

\author{Baihan Lin\\
	Columbia University\\
	New York, NY 10027, USA \\
	\texttt{baihan.lin@columbia.edu} \\
 \And
	Xinxin Zhang \\
	New York University \\
	 New York, NY 10012, USA \\
	\texttt{xz3149@nyu.edu} \\
}

\renewcommand{\headeright}{}
\renewcommand{\undertitle}{}
\renewcommand{\shorttitle}{A Reinforcement Learning Framework for Online Speaker Diarization}

\hypersetup{
pdftitle={A template for the arxiv style},
pdfsubject={q-bio.NC, q-bio.QM},
pdfauthor={Baihan Lin},
pdfkeywords={First keyword, Second keyword, More},
}

\begin{document}
\maketitle

\begin{abstract}

\input{sec_abstract}
\end{abstract}


\input{sec_body}

\bibliographystyle{unsrt}
\bibliography{main}

\end{document}

%% file: sec_abstract.tex
Speaker diarization is a task to label an audio or video recording with the identity of the speaker at each given time stamp. In this work, we propose a novel machine learning framework to conduct real-time multi-speaker diarization and recognition without prior registration and pretraining in a fully online and reinforcement learning setting. Our framework combines embedding extraction, clustering, and resegmentation into the same problem as an online decision-making problem. We discuss practical considerations and advanced techniques such as the offline reinforcement learning, semi-supervision, and domain adaptation to address the challenges of limited training data and out-of-distribution environments. Our approach considers speaker diarization as a fully online learning problem of the speaker recognition task, where the agent receives no pretraining from any training set before deployment, and learns to detect speaker identity on the fly through reward feedbacks. The paradigm of the reinforcement learning approach to speaker diarization presents an adaptive, lightweight, and generalizable system that is useful for multi-user teleconferences, where many people might come and go without extensive pre-registration ahead of time. Lastly, we provide a desktop application that uses our proposed approach as a proof of concept. To the best of our knowledge, this is the first approach to apply a reinforcement learning approach to the speaker diarization task.

%% file: sec_body.tex
\section{Introduction}

Speaker diarization is a crucial task in many real-world applications, such as meeting transcription, call center monitoring, and broadcast news processing. The goal of speaker diarization is to partition an audio or video stream into homogeneous segments, each corresponding to a single speaker, without any prior knowledge of the speakers' identities \cite{tirumala2017speaker,anguera2012speaker}. This task has traditionally been addressed using unsupervised clustering methods \cite{zajic2017speaker,senoussaoui2013study,wang2018speaker}, but recent advances in deep learning have led to the development of more powerful embedding-based approaches \cite{shum2013unsupervised,snyder2018x,wang2018speaker}.

Despite the recent progress, speaker diarization remains a challenging problem, particularly in real-time and online scenarios where new speakers may enter or leave the conversation at any time. In such cases, pre-trained models may not be sufficient, and the system must be able to adapt to new speakers on the fly \cite{lin2020voiceid,lin2020speaker,lin2021speaker}. As in the successful applications to other speech and language tasks \cite{lin2022reinforcement}, the reinforcement learning (RL) has emerged as a promising approach for developing next-generation speaker diarization systems that can learn online and adapt to changing circumstances.

In this paper, we propose a novel RL framework for online speaker diarization that does not require prior registration or pretraining. Our approach combines embedding extraction, clustering, and resegmentation into a single online decision-making problem, where the agent receives feedback in the form of rewards or penalties for each segmentation decision. We demonstrate the effectiveness of our approach using a Q-learning-based diarization agent on a desktop app, and discuss practical considerations for implementing and deploying RL-based speaker diarization systems.

\section{Problem Setting}

\begin{figure}[tb]
\centering
    \includegraphics[width=.5\linewidth]{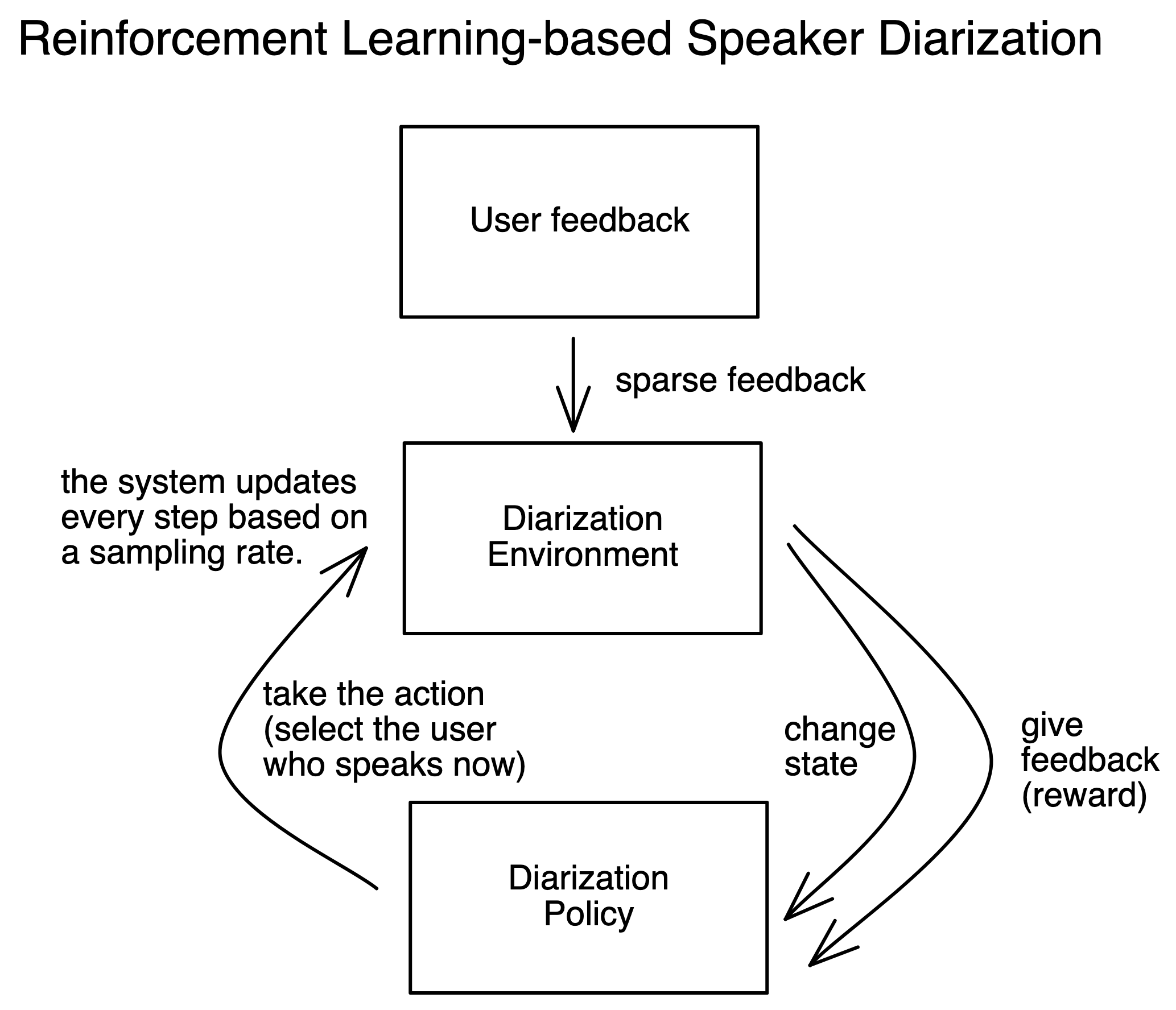}
\caption{The reinforcement learning framework to speaker diarization
}\label{fig:rldiarization}
\end{figure}

 We consider the task of real-time multi-speaker diarization and recognition in a fully online and reinforcement learning setting. Specifically, we aim to develop an intelligent system that can label an audio or video recording with the identity of the speaker at each given time stamp, without prior registration and pretraining. The system must be able to adapt to out-of-distribution environments and learn continually, receiving feedback in the form of rewards from the user.

\subsection{System assumptions}

In this problem setting, we assume that we have no prior knowledge of the number of speakers in the audio stream and the identity of each speaker. The system must be able to detect and recognize speakers on the fly as new speakers enter and leave the conversation, without requiring pre-registration or prior training on large datasets.

We further assume that the system is equipped with a set of contrastive audio embeddings, clustering and resegmentation modules, but it has not been pre-trained on any specific task or dataset. Instead, the system will learn to detect speaker identity through reinforcement learning, receiving feedback in the form of rewards from the user.

The goal of the system is to minimize the error in labeling speaker identity at each time stamp and provide a smooth and accurate experience for the user. By framing speaker diarization as a fully online learning problem, we hope to create a system that is adaptable, lightweight, and able to generalize to new users and environments.

\begin{figure}[tb]
\centering
    \includegraphics[width=\linewidth]{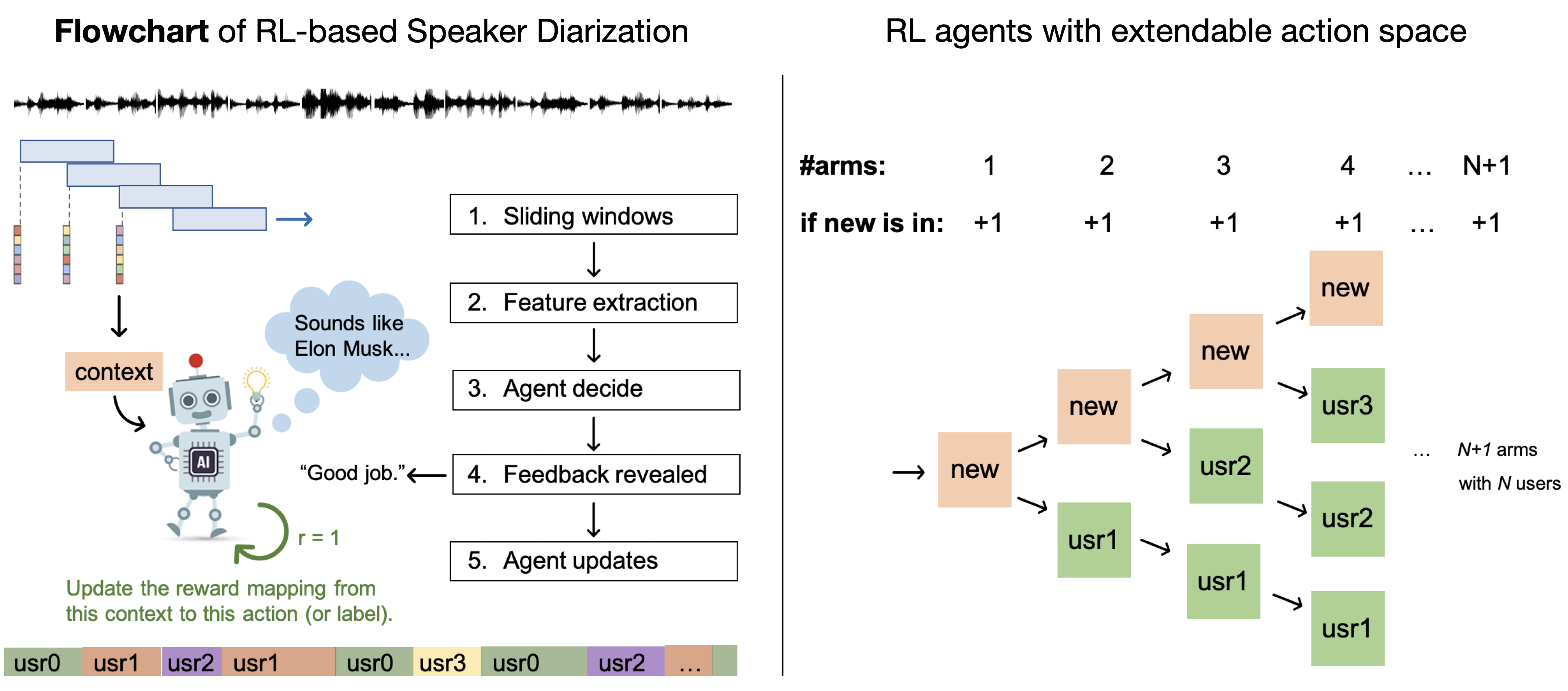}
\caption{(A) The flowchart of an RL-based speaker diarization system, and (B) RL with extendable action space
}\label{fig:flowbandit}
\end{figure}

\subsection{Reinforcement learning formulation}

In the reinforcement learning setting of the online speaker diarization problem, we consider the speaker diarization task as a Markov decision process (MDP). The agent (speaker diarization system) interacts with an environment (audio data) in a sequence of time steps. At each time step, the agent observes the current state of the environment, takes an action, and receives a reward. The objective of the agent is to learn a policy that maximizes the expected sum of future rewards, as outlined in Figure \ref{fig:rldiarization}.
In this problem, the state is the current audio signal segment, and the action is the speaker label assigned to that segment. The reward is the quality of the diarization result, which can be defined in different ways, such as the accuracy of speaker labeling, the precision or recall of the clustering, or the speaker change detection rate. The reward can also be sparse or delayed, which means that the agent may receive little or no feedback for some time steps before receiving a large reward or punishment.

This framework enables the possiblity of incorporating various and multiple streams to \textit{reward signals} into our model learning process during deployment phase. These reward signals can be user-provided feedbacks, by directly asking the users to provide explicit feedback on the speaker diarization performance. For example, after each diarization, the system can ask users to rate the quality of the separation or provide a binary feedback indicating whether the system performed well or not. They can also be model-related feedbacks such as time-based rewards to incentivize the model to improve the separation over time. For example, the system can provide a positive reward for each second of high-quality separation and a negative reward for each second of poor separation. This type of reward signal can encourage the model to focus on separating speakers accurately over the entire duration of the audio stream. The uncertainty of the models can also be taken into account by computing the model's own confidence estimates to confidence-based rewards. Finally, our RL model can combine the above reward signals to create a hybrid reward function. For example, the system can use user-provided feedback to update the model's weights during training and use time-based rewards to encourage the model to separate speakers accurately over time.

The online speaker diarization problem is challenging because it involves an unknown number of speakers, and the speaker identities can change over time. Furthermore, the quality of the diarization result depends not only on the current segment but also on the previous and future segments. Therefore, the speaker diarization system needs to consider long-term dependencies and uncertainty to make effective decisions.

In a RL scenario without prior knowledge of the number of users, the system must dynamically expand its action space as new labels arrive. This problem can be modeled by bandits with infinitely many arms, where new arms are generated when feedback confirms a new addition \cite{berry1997bandit}. For the speaker registration process, we apply the arm expansion process shown in Figure \ref{fig:flowbandit}B, starting from a single arm for the "new" action. At each time step, a segment of speech information is fed into a feature extractor, which can be a neural network or other feature engineering technique, and the agent uses this feature to make a decision on the speaker identity. If a user provides feedback, the agent updates its policy based on this reward, which may be a sparse correction term. Figure \ref{fig:flowbandit}A illustrates the data flow of the system.



\section{Background}

\begin{figure}[tb]
\centering
    \includegraphics[width=\linewidth]{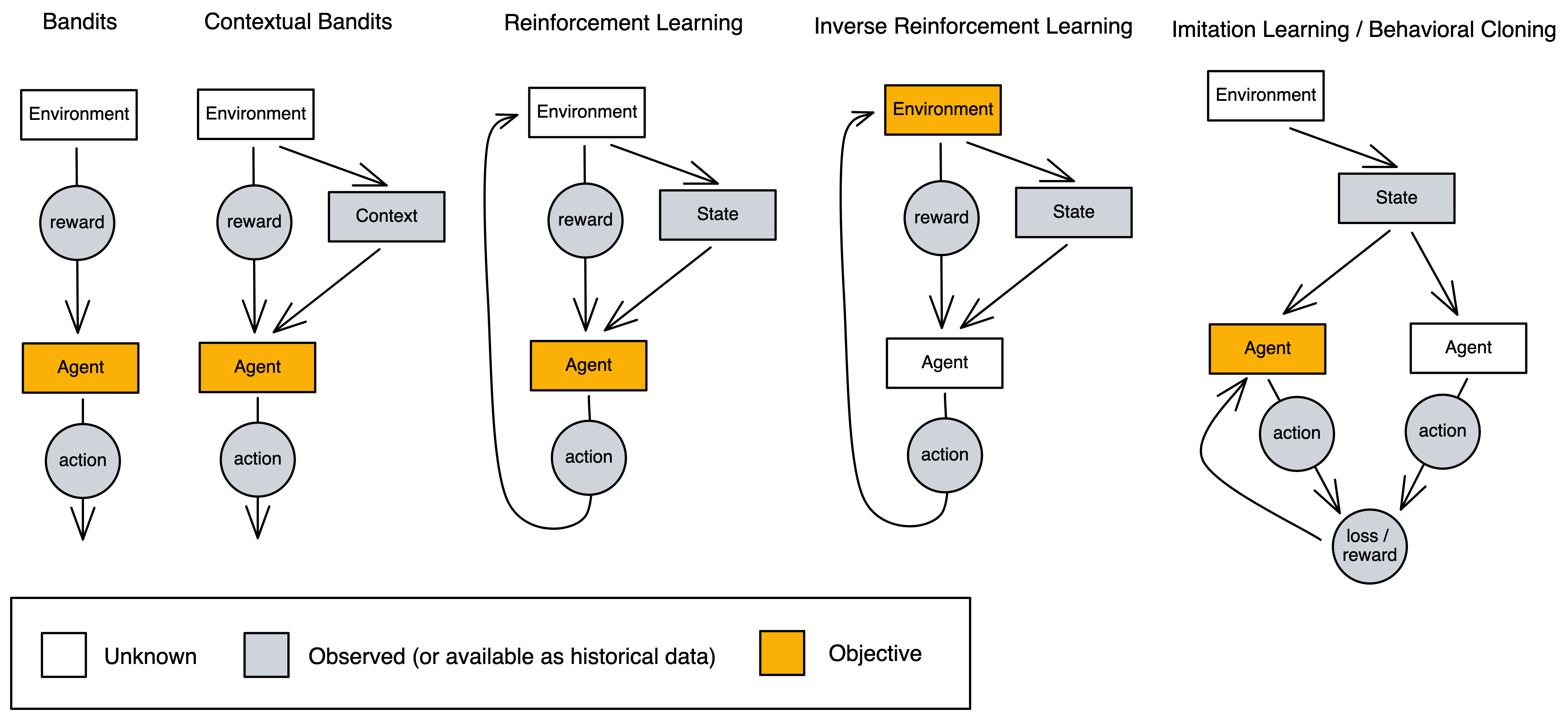}
\caption{Five classes of reinforcement learning problems (from this review on RL in speech applications \cite{lin2022reinforcement})
}\label{fig:rlclasses}
\end{figure}

Reinforcement Learning (RL) is a powerful machine learning paradigm in which an agent learns to make decisions by interacting with the environment. The goal of the agent is to maximize a cumulative reward signal by learning an optimal policy, which maps states to actions. Figure \ref{fig:rlclasses} outlines the problem settings of the five main classes of reinforcement learning approaches, which we believe can all facilitate effective online speaker diarization.

One of the simplest forms of RL is the \textit{Multi-Armed Bandits} (MAB) problem, which consists of a set of $K$ arms, each associated with a stochastic reward distribution. At each time step, the agent chooses an arm to pull, and receives a reward sampled from the corresponding distribution. The goal is to maximize the cumulative reward over a finite number of time steps. The MAB problem is a fundamental trade-off between exploration (trying out new arms to discover the best one) and exploitation (selecting the best arm discovered so far) \cite{LR85,UCB,chapelle2011empirical}.

\textit{Contextual Bandits} (CB) extend the MAB problem by incorporating contextual information, which represents additional features that are specific to each arm. The agent observes a context vector before choosing an action, and the goal is to learn a policy that maps contexts to actions. CB algorithms are widely used in personalized recommendations and online advertising \cite{li2010contextual,langford2008epoch,ChuLRS11}.

\textit{Reinforcement Learning} generalizes MAB and CB to problems with more complex state and action spaces. In RL, the agent observes a state of the environment, takes an action, and receives a reward and the next state. The agent's goal is to learn a policy that maximizes the expected cumulative reward. Some popular RL algorithms include Q-learning \cite{watkins1992q}, SARSA \cite{rummery1994line}, and Actor-Critic \cite{konda1999actor}.

\textit{Inverse Reinforcement Learning} (IRL) is the problem of inferring the reward function given the optimal policy \cite{ng2000algorithms,lin2019split}. IRL has applications in imitation learning, where an agent learns to mimic the behavior of an expert by inferring its underlying reward function \cite{ho2016generative,finn2016guided}.

\textit{Behavioral Cloning }(BC) is a supervised learning method that trains an agent to mimic the expert's actions given the expert's state \cite{pomerleau1991efficient}. BC is simple and computationally efficient, but it suffers from the covariate shift problem, where the distribution of states observed during training differs from the distribution encountered during testing \cite{ross2011reduction}. One exception is the Behavior Cloning with Demonstration Rewards (BCDR) \cite{lin2022ipd,BalakrishnanBMR19}, which is a training procedure for solving this problem by reformulating it as a RL setting. 

\textit{Imitation Learning} (IL) extends BC to address the covariate shift problem by learning a policy that maps states to actions, while leveraging the expert's behavior as a guide \cite{ross2011reduction, abbeel2004apprenticeship}. IL has applications in robotics, where an agent learns to perform complex manipulation tasks by observing an expert \cite{kroemer2010combining}.

More formally, the reinforcement learning defines a class of algorithms for solving problems modeled as Markov decision processes (MDP) \cite{Sutton1998}. An MDP is defined by the tuple $(\mathcal{S}, \mathcal{A}, \mathcal{T}, \mathcal{R}, \gamma)$, where  $\mathcal{S}$ is a set of possible states, $\mathcal{A}$ is a set of actions, $\mathcal{T}$ is a transition function defined as $\mathcal{T}(s, a, s')=\Pr(s'\vert s,a)$, where $s, s'\in \mathcal{S}$ and $a\in \mathcal{A}$, and $\mathcal{R}: \mathcal{S}\times \mathcal{A} \times \mathcal{S}\mapsto \mathbb{R}$ is a reward function, $\gamma$ is a discount factor that decreases the impact of the past reward on current action choice. Typically,  the objective is to maximize the discounted long-term reward, assuming an infinite-horizon decision process.
, i.e. to find a policy function $\pi: \mathcal{S} \mapsto \mathcal{A}$ which specifies the action to take in a  given state, so that the cumulative reward is maximized: $\max_{\pi} \sum_{t=0}^{\infty}\gamma^t \mathcal{R}(s_t,a_t, s_{t+1}).$

\section{Practical Considerations}

In our early work, we have shown a successful proof of concept of a bandit-based speaker diarization system in \cite{lin2021speaker,lin2020speaker,lin2020voiceid,lin2022voice2alliance}. The algorithm which drives the system is a contextual bandit algorithm called the BerlinUCB \cite{lin2020online}. It takes into account the episodicity of the rewards by tackling with a semi-supervised mechanism. Evaluated in the MiniVox benchmark \cite{lin2021speaker}, this one preliminary type of reinforcement learning approach outperforms existing methods by a significant amount. Here we discuss the practical considerations when further extending it to the full spectrum of reinforcement learning problems, beyond the contextual bandit setting.

\subsection{Technical challenges}

There are several innate technical challenges related to applying RL to speaker diarization problems. One of the most significant challenges is dealing with the \textit{sparse feedback problem}. In many real-world applications, the rewards or feedback from the user are sparse, and this can make it difficult to train a model. This can be addressed by using semi-supervised learning methods, such as generative adversarial networks (GANs) or variational autoencoders (VAEs), to generate additional training data from the limited feedback.

The \textit{large state space} of speaker diarization problems is another challenge. To address this, the state space can be reduced by using feature engineering, such as MFCC, and/or dimensionality reduction techniques, such as principal component analysis (PCA) or t-distributed stochastic neighbor embedding (t-SNE).

Another practical consideration is dealing with the problem of \textit{domain adaptation}, which refers to the challenge of adapting the RL model to different domains or environments. One way to address this is by using transfer learning techniques, which involve training the RL model on a source domain and then fine-tuning it on the target domain.

Another consideration is the need for \textit{continual learning}, which is necessary in real-world applications where the environment is constantly changing. To achieve this, online learning algorithms, such as online Q-learning and deep Q-learning, can be used to update the model in real-time as new data becomes available.

Finally, it is important to consider the \textit{scalability of RL algorithms} when dealing with large datasets. To address this, distributed RL algorithms, such as parameter server and actor-critic, can be used to parallelize the training process across multiple machines or devices.

\begin{figure}[tb]
\centering
    \includegraphics[width=.5\linewidth]{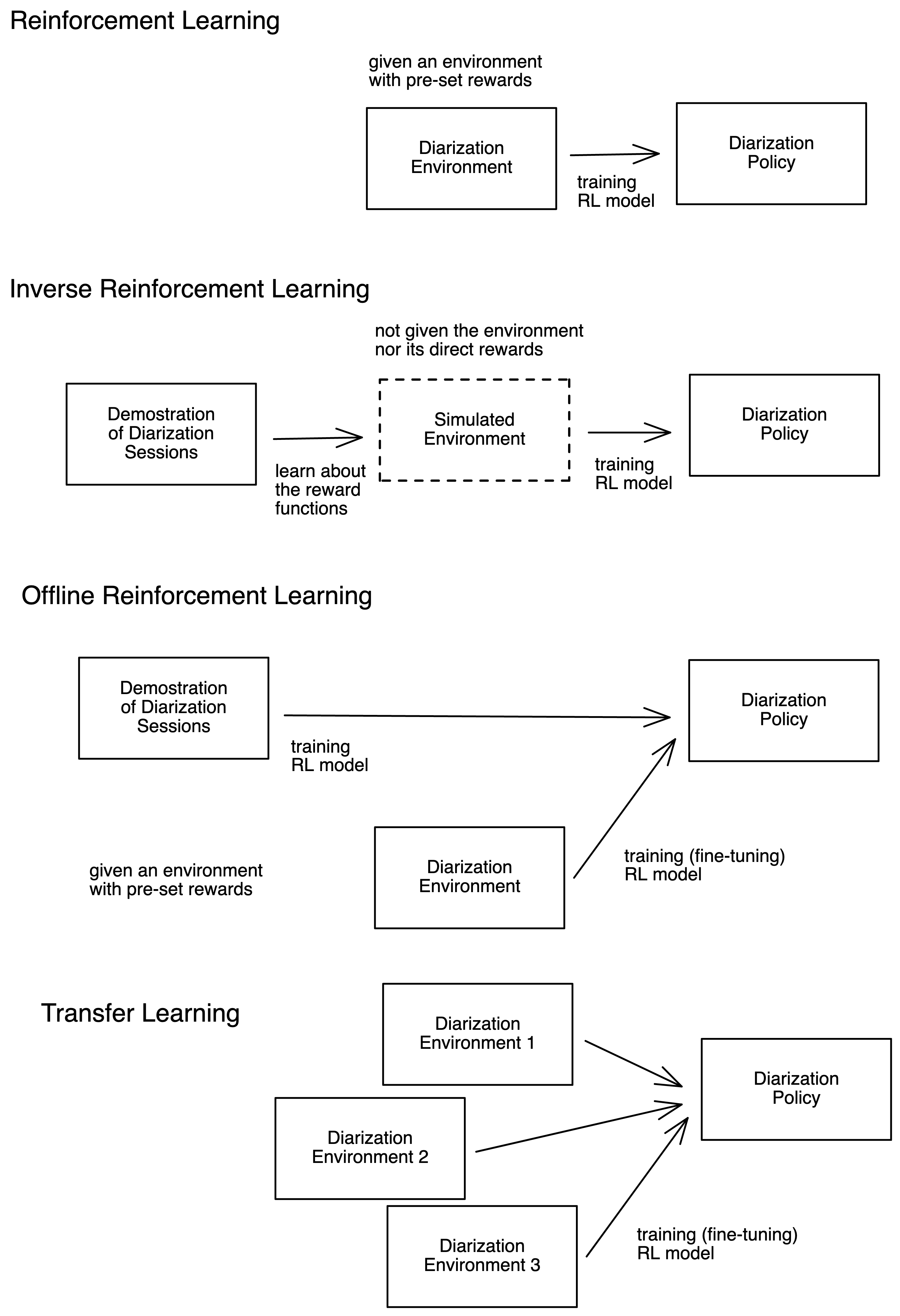}
\caption{Four algorithmic routes of training RL-based diarization systems
}\label{fig:4app}
\end{figure}

\subsection{Algorithmic considerations}

When choosing the right method or algorithms to deploy on speaker diarization domain, there are several advanced strategies worth noting, enabled by our reinforcement learning approach. Figure \ref{fig:4app} summarizes the four possible approaches that it can be useful to improve the training of reinforcement learning for speaker diarization. 

\textit{Deep reinforcement learning} (DRL) has recently been applied to various speaker applications \cite{li2017deep}, along with other \textit{deep learning} based RL or bandits approach \cite{henderson2018deep,collier2018deep}. Deep Q network is a popular method \cite{mnih2013playing}. Compared to traditional reinforcement learning methods, DRL can handle complex high-dimensional input spaces, which is essential for the speaker diarization task since it requires processing audio signals with a high degree of fidelity. However, training deep models requires significant computational resources and can lead to overfitting, which can be problematic when working with limited datasets. Therefore, it is crucial to carefully balance model complexity and training data availability when applying DRL to speaker diarization.

\textit{Inverse reinforcement learning} (IRL) has been proposed as an alternative to traditional reinforcement learning. IRL can learn the underlying reward function that a speaker would receive, based on the observed behavior of the speaker \cite{abbeel2004apprenticeship}. This is a useful technique for the speaker diarization problem since it can allow for the efficient use of historical data, which can help in cases where training data is limited. However, IRL requires more processing time than standard RL techniques and can be more challenging to implement. However, once a reward function is learned, IRL can sometimes provide much more dense reward feedbacks to the diarization system and thus alleviating the aforementioned sparse feedback issue.

\textit{Offline reinforcement learning} (ORL) is a variation of RL where the agent learns from a batch of previously collected data. Some popular methods include Conservative Q-Learning, among others \cite{levine2020offline,kidambi2020morel,kumar2020conservative}. In speaker diarization, ORL can help train the agent on historical audio data, which can then be used to update its speaker recognition policy. ORL can be an attractive alternative to online RL methods since it can be computationally efficient and can handle large-scale datasets. However, ORL assumes that all of the relevant data is available before the agent starts learning, which may not always be feasible in practice.

\textit{Transfer learning} is a method for reusing a previously trained model to improve performance on a new but related task \cite{pan2009survey}. Transfer learning is useful in the speaker diarization problem since it allows the agent to use a previously trained model to recognize speaker identity for a new speaker without the need to retrain the model from scratch. For instance, the speaker diarization system trained on adult speech corpus can be helpful to kick start a system for kids.  Transfer learning can be applied to both offline and online RL methods and can significantly reduce the amount of data required to train a model. 

When applying these algorithms to speaker diarization problems, it is essential to pay attention to the challenges unique to the speaker diarization task. Specifically, since speaker diarization involves processing audio signals in real-time, there is a need for efficient and lightweight algorithms that can process audio signals quickly. Furthermore, speaker diarization requires the recognition of multiple speakers, which can make it challenging to provide reward signals for RL agents. Careful selection of the RL algorithm and tuning of its hyperparameters is necessary to address these challenges and achieve high performance in speaker diarization tasks.


\begin{figure}[tb]
\centering
    \frame{\includegraphics[width=.48\linewidth]{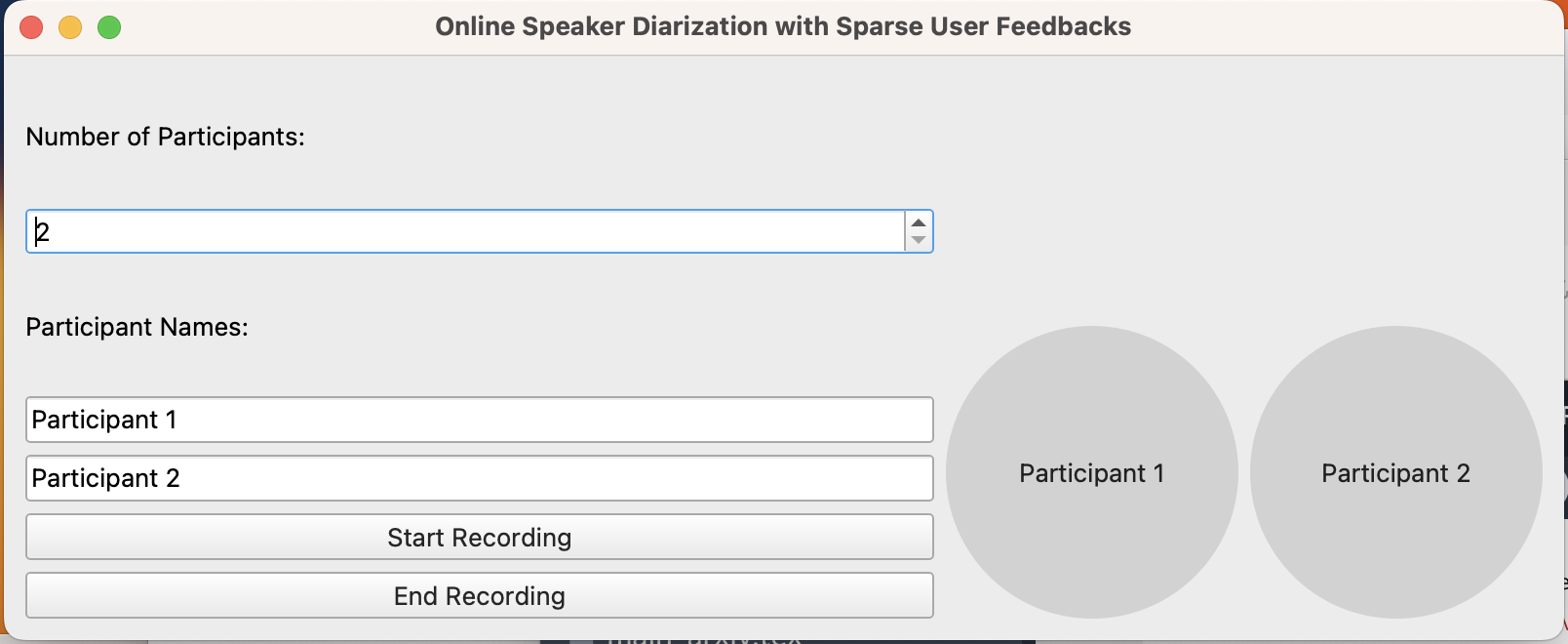}} \hfill
    \frame{\includegraphics[width=.48\linewidth]{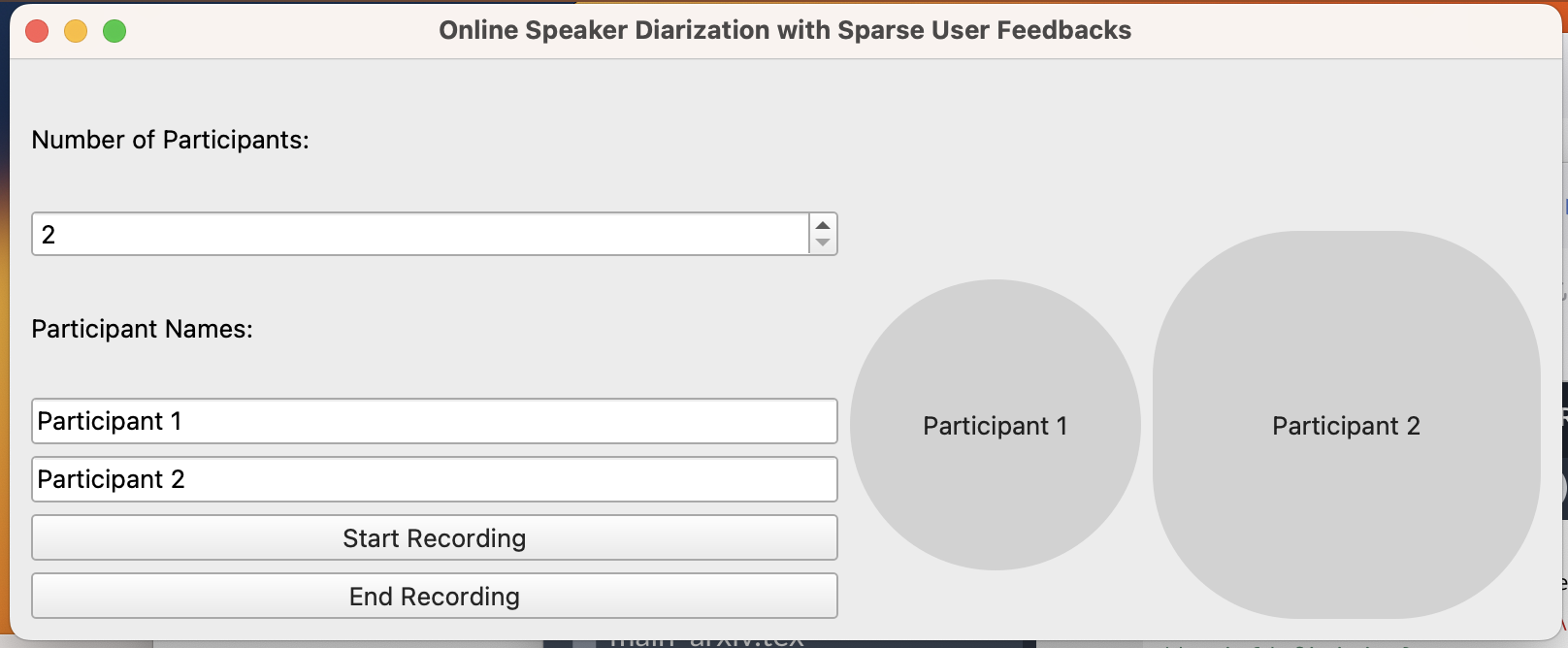}}
\par\caption{Screenshots of the Desktop App as a proof of concept to the user scenarios.}\label{fig:app}
\end{figure}

\section{Demonstration System}

We have developed a desktop application to demonstrate our proposed online speaker diarization system using reinforcement learning. The application is built using Python and the PyQt5 framework for the graphical user interface. The reinforcement learning algorithm is implemented using the Q-learning method with an epsilon-greedy exploration strategy. The audio signal processing is handled by the librosa library, which provides functions for feature extraction such as Mel-frequency cepstral coefficients (MFCCs).

The user interface of the application allows the user to input the number of participants in the audio stream and their names (Figure \ref{fig:app}). The system then generates buttons for each participant, which the user can click to indicate which participant is currently speaking. The Q-learning algorithm updates the policy for speaker diarization based on the user feedback. In our earlier web-based demonstration using bandits \cite{lin2020voiceid}, the system also has the option to provide a live visualization of the audio stream, showing the audio waveform and the currently active speaker. In \cite{lin2021speaker}, we have evaluated the system using a dataset of multi-speaker audio recordings and compared its performance to that of a traditional clustering-based speaker diarization method. Our results show that the Q-learning-based approach outperforms the traditional method in terms of accuracy and adaptability to changing speaker configurations.

Overall, our demonstration system showcases the potential of using online reinforcement learning for real-time speaker diarization without requiring pretraining or registration of users. The system is adaptable to different speaker configurations and can learn from user feedback to improve its performance over time.



\section{Conclusion}

In this paper, we proposed a novel reinforcement learning framework for online speaker diarization. Our framework provides a solution for real-time multi-speaker recognition and diarization without requiring prior registration and pre-training. The proposed framework combines the embedding extraction, clustering, and re-segmentation into a single online decision-making problem.

We discussed practical considerations in applying reinforcement learning to speaker diarization problems, including the limitations of data availability, the challenge of sparse feedback, and the need for domain adaptation. We also reviewed various reinforcement learning algorithms, including multi-armed bandits, contextual bandits, reinforcement learning, inverse reinforcement learning, behavioral cloning, and imitation learning, and discussed their potential applications and limitations in the context of speaker diarization. To intrigue the audience into this growing field, we provided a demonstration system which illustrates the potential of our reinforcement learning framework in a desktop application that allows users to perform real-time speaker recognition and diarization. 

In summary, our reinforcement learning framework offers a promising approach to online speaker diarization, which has the potential to be more adaptive, lightweight, and generalizable to new users. With further development and research, we hope that our framework can become a useful tool in various real-world applications, including multi-user teleconferencing, call center management, and security monitoring.